# Strange particles production in relativistic nucleus-nucleus collisions at RHIC and LHC energy regions *

REN Xiao-Wen(任小文)[1], FENG Sheng-Qin(冯笙琴)[1,2,3;)1] YUAN Xian-Bao（袁显宝）[1]

[1] College of Science, China Three Gorges University，Yichang 443002, China

[2] Key Laboratory of Quark and Lepton Physics (Huazhong Normal University), Ministry of Education，Wuhan 430079，China

[3] School of Physics and Technology, Wuhan University, Wuhan 430072, China

**Abstract:** PACIAE, a parton and hadron cascade model, is utilized to systematically investigate strange particle production and strangeness enhancement in Au+Au collision and in Pb+Pb collision with the $\sqrt{s_{NN}}$ 200GeV at RHIC and 2.76TeV at LHC, respectively. The experimental results at different centrality, which come from the STAR collaboration and the ALICE collaboration, are well described by the PACIAE model. This may represent the importance of the parton and hadron rescatterings, as well as the reduction mechanism of strange quark suppression, implemented in the PACIAE model.

## 1 Introduction

Relativistic heavy ion collisions aim to create the Quark Gluon Plasma (QGP），a unique state of matter where quarks and gluons can move freely over large volumes in comparison to the typical size of a hadron. Strange particle production and strangeness enhancement in relativistic nucleus-nucleus collisions relative to *pp* collisions at the same energy has been proposed as a signature of the QGP formation in relativistic heavy ion collisions [1]. This is based on the principle that the threshold energy of strange particle production in QGP is higher than that in hadronic matter. Therefore, the yield of strangeness is a good probe of QGP.

Recently, the ALICE collaboration working at LHC (Large Hadron Collider) reported the new data of strange particles production in Pb+Pb collision at $\sqrt{s_{NN}}$ =2.76TeV [2], this is a new

* Supported by National Natural Science Foundation of China (11247021，10975091), Excellent Youth Foundation of Hubei Scientific Committee (2006ABB036) and Key Laboratory foundation of Quark and Lepton Physics (Hua-Zhong Normal University)(QLPL201101)

 1) Corresponding author: fengsq@ctgu.edu.cn

energy regime in relativistic nucleus-nucleus collisions. The STAR collaboration [3] working at Relativistic Heavy Ion Collider (RHIC) published the results of strange particles production in Au+Au at $\sqrt{s_{NN}}$ =200GeV. Based on PYTHIA, the PACIAE [4, 5] model is used to analyze the strange particles production in relativistic nuclear-nuclear collision.

In the LUND string fragmentation scheme [6], the suppression of *s* quark pair production compared with *u* (*d*) pair production (the parameter *parj*(2) in PYTHIA or $\lambda$ denoted later) was assumed to be fixed. However, later experiments [7] have shown that this suppression decreases with increasing reaction energy. In Ref. [8], the reduction mechanism of the strange quark suppression has been introduced in the LUCIAE model. By this mechanism, they described the strangeness enhancement in *pp*, *p* + *A*, and *A* + *A* collisions at CERN Super Proton Synchrotron (SPS) energies successfully [8, 9].

In order to study the strange particles production and strangeness enhancement in relativistic *pp* collisions, we introduced the reduction mechanism of the strange quark suppression in the PACIAE model [10, 11]. In this paper, we will use the reduction mechanism of the strange quark suppression to investigate the relativistic nucleus-nucleus collisions at the RHIC and Large Hadron Collider (LHC) energy regions.

This paper is organized as follows. In Sec.2, we give a brief review of the PACIAE model and the reduction mechanism of strangeness quark suppression. In Sec.3, we use the parton and hadron cascade model PACIAE to analyze systematically the strangeness production in Au + Au collisions at RHIC energies and Pb + Pb collisions at LHC energies. Sec.4 gives a summary and conclusion.

## 2  The new modified PACIAE models

Based on PYTHIA [12], the parton and hadron cascade model PACIAE is a model can study nucleus-nucleus collisions. The PACIAE model has four stages such as: the parton initialization, parton evolution (rescattering), hadronization, and hadron evolution (rescattering).

⑴ The parton initialization: Based on the collision geometry, a nucleus-nucleus collision is decomposed into the nucleon-nucleon collisions in the PACIAE model. A nucleon-nucleon (NN) collision is decrypted by the PYTHIA model, by which a hadron- hadron (h-h) collision is decomposed into the parton-parton collisions, with the string fragmentation (hadronization) process switched-off. This parton configuration is regarded as QGP formed in the initial state of h-h collision. Then one can obtain a parton configuration composed of quarks, anti-quarks, and gluons, besides a few hadronic remnants for a h-h collision after diquarks (anti-diquarks) being split randomly into quarks (anti-quarks).

(2) The parton evolution (rescattering): The rescattering among partons in QGP is taken into account by the 2→2 LO-pQCD parton-parton cross sections [13]. One can calculates the differential cross section of a subprocess $ij \to kl$ by

$$\frac{d\sigma_{ij \to kl}}{d\hat{t}} = K \frac{\pi \alpha_s^2}{\hat{s}} \sum_{ij \to kl} \qquad (1)$$

Where $\alpha_s = 0.47$ stands for the effective strong coupling constant, $\hat{s}$, $\hat{t}$, as well as $\hat{u}$ refer to the Mandelstam variables and the factor $K$ is introduced considering the higher order and the non-perturbative corrections. Taking the process q1q2 →q1q2 as an example，we can get

$$\sum_{q_1 q_2 \to q_1 q_2} = \frac{4}{9} \frac{\hat{s}^2 + \hat{u}^2}{\hat{t}^2} \ . \qquad (2)$$

By introducing the parton color screen mass $\mu = 0.63$ GeV, one can get

$$\sum_{q_1 q_2 \to q_1 q_2} = \frac{4}{9} \frac{\hat{s}^2 + \hat{u}^2}{\left(\hat{t}^2 - \mu^2\right)^2} \qquad (3)$$

the total cross section of $i + j$ parton collision is then

$$\sigma_{ij}(\hat{s}) = \sum_{k,l} \int_{-\hat{s}}^{0} d\hat{t} \frac{d\sigma_{ij \to kl}}{d\hat{t}}. \qquad (4)$$

The total and differential cross sections above the parton evolution (rescattering) can be simulated by the Monte Carlo method until all parton-parton collisions are exhausted (partonic freeze-out).

(3) The hadronization: In the hadronization stage, the partonic matter is hadronized by the Lund string fragmentation model [12] or by the Monte Carlo coalescence model.

(4) The hadron evolution (rescattering): The rescatterings among $\pi, K, p, n, \rho(\omega), \Delta, \Lambda, \Sigma,$ $\Xi, \Omega, J/\Psi$ and their antiparticles are taken into account for the moment. The isospin averaged parametrization formula [14, 15] is assumed for cross section of h-h collisions. In addition, some assumed constant total cross sections $\sigma_{tot}^{NN} = 40 mb$, $\sigma_{tot}^{\pi N} = 25 mb$, $\sigma_{tot}^{KN} = 20 mb$, $\sigma_{tot}^{\pi\pi} = 10 mb$ and ratio of inelastic to total cross section (0.85) are offered as another option.

In the LUND string fragmentation regime, the $q\bar{q}$ pair with mass $m$ and transverse momentum $p_t$ may be created quantum mechanically in one point and then tunnel out to the classically allowed region. This tunneling probability is given by

$$\exp\left(-\frac{\pi m^2}{\kappa}\right)\exp\left(-\frac{\pi p_T^2}{\kappa}\right) \quad (6)$$

where the string tension $\kappa \approx 1\text{GeV/fm} \approx 0.2\text{GeV}^2$ [6, 12]. This probability implies a suppression of strange quark production u:d:s:c≈1:1:0.3:$10^{-11}$. The charm and heavier quarks are not expected to be produced in the soft string fragmentation process provided no charm and heavier quarks appearing in the string, but only in the hard process or as part of the initial- and final-state QCD radiations. The higher $p_T$ quark pair is expected to be created provided the string tension is large.

A mechanism of the increase of effective string tension and hence the reduction of strange quark suppression was introduced in Ref.[8]. In that paper, it was assumed that the effective string tension increases with the increasing of the number and the hardening of gluons in the string according to

$$\kappa^{eff} = \kappa_0 (1-\xi)^{-\alpha} \quad (7)$$

$$\xi = \frac{\ln\left(\frac{k_{T\max}^2}{s_0}\right)}{\ln\left(\frac{s}{s_0}\right) + \sum_{j=2}^{n-1}\ln\left(\frac{k_{Tj}^2}{s_0}\right)} \quad (8)$$

where $\kappa_0$ is the string tension of the pure $q\bar{q}$ string assumed to be ~1GeV/fm. The gluons in the multi-gluon string are ordered from *2* to *n-1* because of two ends of string are quark and anti-quark. $k_{Tj}$ is the transverse momentum of gluon $j$ with $k^2_{Tj} \geq s_0$ and $k_{Tmax}$ is the largest transverse momentum among the gluons. The parameters *a=3.5GeV* and *s₀=0.8 GeV* are determined by fitting h-h collision data. It should be mentioned that the Eq. (8) represents the deviation scale of the multigluon string from that of the pure string.

In the PYTHIA model, there are four adjustable parameters:

- parj(1) is the suppression of diquark-antidiquark pair production compared with quark-antiquark production,
- parj(2) is the suppression of *s* quark pair production compared with *u* or *d* pair production,

- parj(3) is the extra suppression of strange diquark production compared with the normal suppression of strange quark,

- parj(21) corresponds to the width $\sigma$ in the Gaussian $p_x$ and $p_y$ transverse momentum distributions for primary hadrons.

They can be found easily in the PACIAE model parameters, similarly. They are able to be tuned to reduce the strange quark suppression and to change the width of its $p_T$ distribution.

If $\lambda$ denotes parj(2) then by Eq.(6), we can obtain

$$\lambda_2 = \lambda_1^{\frac{\kappa_1^{eff}}{\kappa_2^{eff}}} \qquad (9)$$

$$\sigma_2 = \sigma_1 \left(\frac{\kappa_2^{eff}}{\kappa_1^{eff}}\right)^{1/2} \qquad (10)$$

where the same subscript represent the same string, $k_{eff1}$ ($k_{eff2}$) is effective string tension in Eq.(7). It is not hard to prove that Eq. (9) is also valid for *parj*(1) and *parj*(3).

The reduction mechanism of the strange quark suppression has been included in the PACIAE model (tune parameter kjp22=1). One might first tune the parameters *parj*(1), *parj*(2), and *parj*(3) (assuming parameter parj(21) is a constant in this paper) to fit the strangeness production data in a given nuclear collision system at a given energy. The resulting *parj*(1), *parj*(2), and *parj*(3) can be used to predict the strangeness production in the same reaction system at different energies, even in different reaction systems.

## 3  Calculations and results

Default values of the model parameters in PACIAE were given according to the experimental measurement and/or the physical argument [4]. However, in the specific calculations, a few sensitive parameters, such as $K$, $\beta$ and $\Delta t$ in this paper, should be tuned to a datum of the global measurable, for instance, the charged multiplicity or the charged particle rapidity density at mid-rapidity. The fitted parameters were then used to investigate the other physical observables.

In this paper, for the parameter $K$, has mentioned in Sec. 2, is introduced to consider the higher order and non-perturbative corrections for LO-pQCD parton–parton differential cross section, the parameter $\beta$ is a fact from LUND string fragmentation function, and the parameter $\Delta t$ is the least time interval of two distinguishably consecutive collisions in the parton initiation stage.

The parameters, $K$, $\beta$ and $\Delta t$ should be tuned firstly to suitable value, making the charged multiplicity or the charged particle rapidity density at mid-rapidity of each centrality bin calculated by PACIAE model fit to the global measurable. Then we tuned the parameters parj(1), parj(2), parj(3) to a datum of the experiment results, such as the yields of strange particles. The fitted parameters were then used to investigate the other physical observables, for instance transverse momentum distribution in this paper.

We first globally tuned the parameters *parj*(1), *parj*(2), and *parj*(3) in PACIAE simulations to fit the strangeness production data from Au+Au collisions at $\sqrt{s_{NN}}$ = 200GeV. The yields of strange particles calculated by PACIAE compared to the STAR results are showed in Table I. From the Table I, we know that the data of strange particles yields from the STAR collaboration are better reproduced by PACIAE with the reduction mechanism of the strange quark suppression.

***Table 1*** *Strange particle rapidity densities at midrapidity ($|y|<0.5$) in relativistic Au+Au collisions at $\sqrt{s_{NN}}$ = 200GeV. The STAR data were taken from Ref. [2].*

| 中心度 | dN/dy | | | | | |
|---|---|---|---|---|---|---|
| | $K_s^0$ | | $\Lambda$ | | $\overline{\Lambda}$ | |
| | STAR | PACIAE | STAR | PACIAE | STAR | PACIAE |
| 0~5% | 43.5±2.4 | 43.66 | 14.8±1. | 16.04 | 11.7±0.9 | 12.01 |
| 10%~20% | 27.8±1.4 | 27.57 | 9.16±0.89 | 9.84 | 7.27±0.55 | 7.30 |
| 20%~40% | 16.5±0.83 | 16.18 | 5.70±0.55 | 5.74 | 4.53±0.34 | 4.21 |
| 40%~60% | 7.26±0.49 | 6.91 | 2.38±0.23 | 2.32 | 1.82±0.14 | 1.83 |
| 60%~80% | 2.14±0.19 | 2.09 | 0.71±0.07 | 0.68 | 0.55±0.04 | 0.54 |

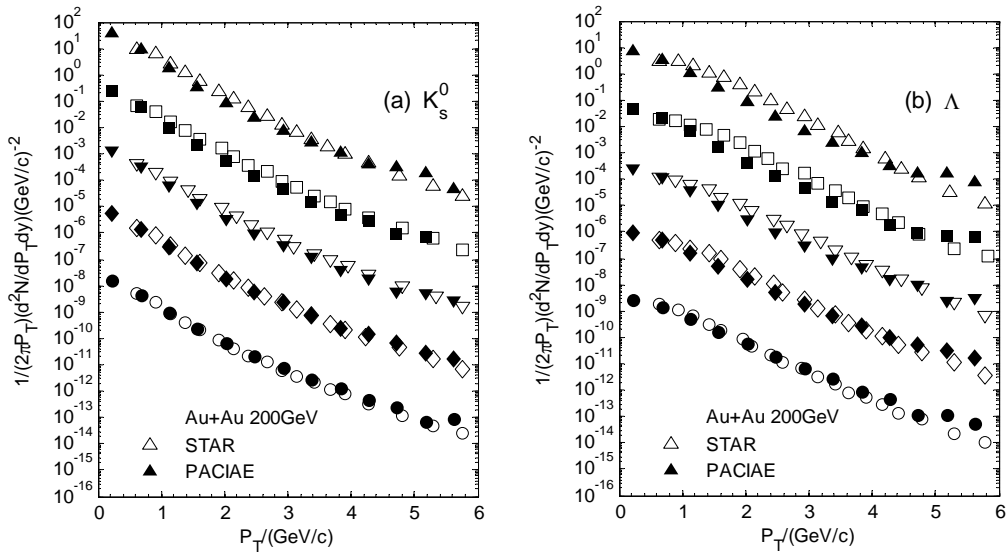

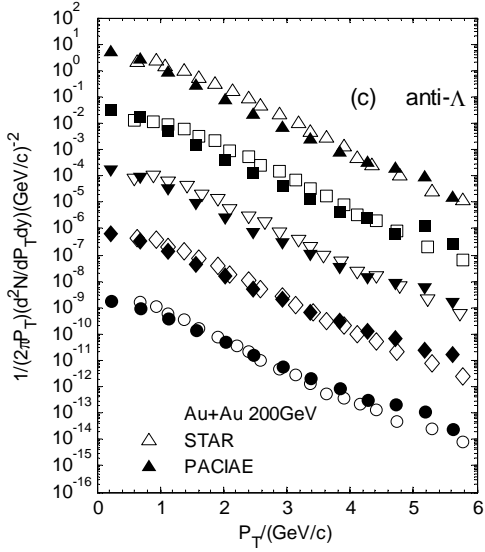

FIG. 1. the transverse momentum distributions of strange particles in relativistic Au + Au collisions at $\sqrt{s_{NN}}$ = 200GeV. Panels (a), (b), and (c) are for $K_S^0$, $\Lambda$, and $\overline{\Lambda}$, respectively. The STAR results are given from Ref. [3]

The transverse momentum spectra ($0 < p_T < 6$ GeV/c) of the strange particles in the relativistic Au+Au collisions at $\sqrt{s_{NN}}$ = 200GeV are showed in Fig. 1, and the corresponding STAR data are from Ref. [3]. The panels (a), (b), and (c) in Fig.1 are for $K_S^0$, $\Lambda$, and $\overline{\Lambda}$, respectively. In order to present the results at different centralities in an identical figure, we multiply the data at 0%~5%, 10%~20%, 20%~40%, 40%~60% and 60%~80% by $10^0$, $10^2$, $10^4$, $10^6$ and $10^8$, respectively. These results indicate that the STAR data of strange particles transverse momentum spectrum is well described by the PACIAE.

Furthermore, comparing with the data of charged particle rapidity density at mid-rapidity [16], we tuned the parameters of $K$, $\beta$ and $\Delta t$ to fit the data at each of centrality bin. We studied the results of strange particles productions in Pb + Pb collisions at $\sqrt{s_{NN}}$ = 2.76TeV [2] by the PACIAE. The transverse momentum spectra ($0 < p_T < 6$) of $K_S^0$ and $\Lambda$ particle at ALICE are showed in Fig. 2 (a, b), respectively. In order to present the results at different centralities in an identical figure, we multiply the data at 0%~10%, 10%~20%, 20%~40%, 40%~60% and 60%~80% by 4.0, $2.0\times10^2$, $1.5\times10^4$, $10^6$ and $10^8$, respectively. From Fig. 2, we can see that the ALICE data of strange particles transverse momentum spectrum can also be well described by the PACIAE.

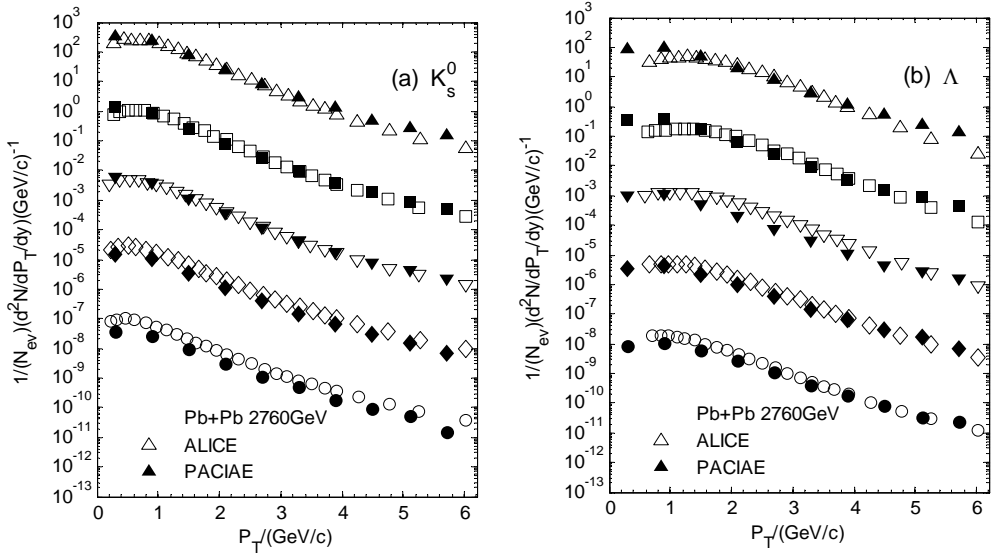

FIG. 2. The transverse momentum distributions of strange particles in relativistic Pb+Pb collisions at $\sqrt{s_{NN}}$ = 2.76TeV. Panels (a), and (b) are for $K_S^0$, and $\Lambda$, respectively. The ALICE data are from Ref. [2].

## 4 CONCLUSIONS

In summary, we have utilized the PACIAE model with the reduction mechanism of the strange quark suppression to analyze strange particles production in relativistic Au+Au collisions at $\sqrt{s_{NN}}$ =200GeV and Pb + Pb at $\sqrt{s_{NN}}$ = 2.76TeV. The PACIAE results of the strange particles rapidity densities at midrapidity($|y|<0.5$) and transverse momentum distributions ($0<p_T<6$) are compared with STAR data and ALICE data, respectively. We find that in general the experiment can be described equally well by the PACIAE model with the reduction mechanism of the strange quark suppression.

In Ref. [4,10], we demonstrated that the effect of the reduction mechanism of strange quark suppression and the parton and hadron re-scatterings introduced in the modified PACIAE model are reasonable to the *pp* collision at RHIC and LHC energy region. Here it is found that the effect of the reduction mechanism of strange quark suppression and the parton and hadron re-scatterings are also reasonable to nuclear-nuclear collisions at RHIC and LHC energy region.

In this paper, we used the PACIAE to study only some strangeness productions such as $\kappa_S^0$, $\Lambda$, and $\bar{\Lambda}$ whose yields are larger than the other strangeness particles such as $\Xi^-$, $\Omega^-$ and theirs antiparticles. In order to further study on the strangeness production mechanism, we also need to check the strangeness production mechanisms of $\Xi^-$, $\Omega^-$ and theirs antiparticles. We will further do this research in our next work.

*The authors are indebted to Prof. Sa Ben-Hao for his valuable discussions and very helpful suggestions.*